\documentclass[namedreferences]{SolarPhysics}
\usepackage[hyperref,optionalrh,natbib]{spr-sola-addons} 
\usepackage{breakurl}
\usepackage{graphicx}                    
\usepackage{amssymb}                     
\usepackage[usenames]{color}             
\usepackage{url}                         
\usepackage{sidecap}                     
\usepackage{lscape}

\newcommand{\BE}{\begin{equation}}
\newcommand{\EE}{\end{equation}}
\newcommand{\BA}{\begin{eqnarray}}
\newcommand{\EA}{\end{eqnarray}}
 \newcommand{\fig}[1]{Figure~\ref{fig:#1}}
 \newcommand{\figs}[2]{Figures~\ref{fig:#1} and \ref{fig:#2}}
 
 \newcommand{\sect}[1]{Section~\ref{sec:#1}}
 \newcommand{\sects}[2]{Sections~\ref{sec:#1} and \ref{sec:#2}}

\newcommand{\degree}{^{\circ}}

\newcommand{\Rsun}{R$_{\odot}$}  

\newcommand{\eg}{\textit{e.g.}}
\newcommand{\ie}{\textit{i.e.}}
\newcommand{\insitu}{{\it in-situ}}

\definecolor{durazno}{rgb}{1.0,0.35,0.0}
\definecolor{anana}  {rgb}{0.0,0.50,1.0}
\definecolor{violeta}{rgb}{1.0,0.00,1.0}

\begin{document}
\begin{article}
\begin{opening}
\title{How Can Active Region Plasma Escape into the Solar Wind from below a Closed Helmet Streamer?}

\author{C.H.~\surname{Mandrini}$^{1,2}$ \sep 
F.A.~\surname{Nuevo}$^{1,2}$ \sep 
A.M.~\surname{V\'asquez}$^{1,2}$\sep
P.~\surname{D\'emoulin}$^{3}$ \sep 
L.~\surname{van~Driel-Gesztelyi}$^{3,4,5}$ \sep
D.~\surname{Baker}$^{4}$ \sep 
J.L.~\surname{Culhane}$^{4}$ \sep 
G.D.~\surname{Cristiani}$^{1,2}$  \sep
M.~\surname{Pick}$^{3}$}

\runningauthor{C.H. Mandrini \textit{et al.}}
\runningtitle{AR Plasma and the Slow Solar Wind}

\institute{$^{1}$ Instituto de Astronom\'\i a y F\'\i sica del Espacio (IAFE), CONICET-UBA, Buenos Aires, Argentina\\
             $^{2}$ Facultad de Ciencias Exactas y Naturales (FCEN), UBA, Buenos Aires, Argentina\\
             email: \url{mandrini@iafe.uba.ar}\\
             $^{3}$ Observatoire de Paris, LESIA, UMR 8109 (CNRS), F-92195 Meudon Principal Cedex, France \\   
             $^{4}$ UCL-Mullard Space Science Laboratory, Holmbury St. Mary, Dorking, Surrey, RH5 6NT, UK\\           
             $^{5}$ Konkoly Observatory, Research Centre for Astronomy and Earth Sciences, Hungarian Academy of Sciences, 
             Budapest, Hungary}
             
\begin{abstract}
Recent studies show that active-region (AR) upflowing plasma, observed by the {\it EUV-Imaging Spectrometer}
(EIS),
onboard {\it Hinode}, can gain access to open field-lines and be released into the solar wind (SW)
{\it via} magnetic-interchange reconnection at magnetic null-points in pseudo-streamer configurations. When only one bipolar AR is present on the Sun and it is fully covered by the separatrix of a streamer, such as AR 10978 in December 2007, it seems unlikely that the upflowing AR plasma can find its way into the slow SW. However, signatures of plasma with AR composition have been found at 1 AU by \inlinecite{Culhane14} apparently originating from the West of AR 10978. 
We present a detailed topology analysis of AR 10978 and the surrounding large-scale corona based on a potential-field source-surface (PFSS) model. Our study shows that it is possible for the AR plasma to get around the streamer separatrix and be released  into the SW {\it via} magnetic reconnection, occurring in at least two main steps. We analyse data from the {\it Nan\c cay Radioheliograph} (NRH) searching for evidence of the chain of magnetic reconnections proposed. We find a noise storm above the AR and several varying sources at 150.9 MHz. Their locations suggest that they could be associated with particles accelerated during the first-step reconnection process and at a null point well outside of the AR. However, we find no evidence of the second-step reconnection in the radio data. Our results demonstrate that even when it appears highly improbable for the AR plasma to reach the SW, indirect channels involving a sequence of reconnections can make it possible.
\end{abstract}
%
\end{opening}

\section{Introduction}
\label{sec:intro}

Parker's seminal work (\opencite{Parker58}, \citeyear{Parker63}) postulated that
the million-degree high-conductivity solar corona must be in continuous
expansion and laid the foundations of a new branch of solar-heliospheric
physics: the study of the origin and characteristics of the solar wind (SW). 
{The steady faster component of the SW originates
from  large coronal holes (CHs),
while small CHs produce a slower wind 
(\opencite{Miralles01}, \citeyear{Miralles04}; \opencite{Neugebauer02})}.
Additionally, a much more variable and slow component of the SW
is frequently associated with coronal streamers. In the case of bipolar
streamers, this slow wind is usually observed at 1 AU around the heliospheric
current sheet (HCS). Most importantly, while the fast steady wind shows a
stable chemical composition, with values similar to those at the photospheric
level (\opencite{vonSteiger97}, \citeyear{vonSteiger01}; \opencite{Zurbuchen99},
\citeyear{Zurbuchen02}), the slow wind is characterized by variable
abundances, with values typical of closed-field active regions. Such abundances are mainly dependent on the first ionization potential (FIP) of the element considered, with the high-FIP elements being typically enhanced 
compared to low-FIP elements in a wide range that goes from two to ten, with typical values for ARs between three 
to five (\eg\ \opencite{Feldman03}; \opencite{Zurbuchen06};
\opencite{Brooks11}). However, there is still no clear consensus of opinion 
from where the slow unsteady wind originates. 

The scenarios proposed for the origin of the slow wind can be divided
into five categories:
   \begin{enumerate}
 \item[i)] The slow wind originates from the boundary of CHs
(\opencite{Suess79}; \opencite{Wang90}; \opencite{Cranmer07}). In these models, the terminal SW speed
is inversely related to the expansion factor of the open field lines, which in turn
depends on the height of heat deposition by magnetohydrodynamic (MHD) waves.
The relationship is such that for lower deposition
heights the terminal SW speed is smaller (\opencite{Holzer80}). However, these
models cannot explain the variable FIP-biased composition of the slow wind. 

 \item[ii)] The interchange reconnection  model (\opencite{Fisk98}; \opencite{Fisk03}; \opencite{Fisk09})
suggests that over quiet-Sun areas closed loops and open-field lines are both
present and the slow wind originates from magnetic reconnection between
them, induced by a continuously changing magnetic carpet. This model
naturally explains the dynamic properties of the slow wind; however,
\inlinecite{Antiochos07} questioned whether mixing of open and closed fields
over quiet-Sun regions is possible, based on Lorentz-force-balance
considerations.

 \item[iii)] The slow wind originates at streamer tops. Closed flux may expand
and open up, and- or interchange reconnection can take place at those locations between open
and closed-field lines (\opencite{Wang00}). In spite of extensive supporting
observational evidence, the narrowness of the latitudinal extent of the
reconnecting region implies that other mechanisms should also be at work
in order to match the much broader latitudinal extent over which the slow wind is
observed (\opencite{McComas98}).

 \item[iv)] The most recent so-called separatrix-web (S-Web) model of slow-wind
formation was proposed jointly by
\inlinecite{Antiochos11}, \inlinecite{Titov11}, and \inlinecite{Linker11} and
further developed by \inlinecite{Antiochos12}. Their model is based on the
``uniqueness conjecture'', discussed by \inlinecite{Antiochos07}, which states
that ``any unipolar region on the photosphere can contain at most one CH''. This implies that 
multiple CHs are connected. The connecting
corridors are thought to be thin, occasionally infinitely thin. 
As these corridors represent drastic change of magnetic
connectivity in the corona, they are, in fact, quasi-separatrix layers (QSLs:
\opencite{Demoulin96a}), which might include separatrices (\opencite{Masson09}). 
The slow wind is suggested to originate along these corridors as they
continuously open and close down by reconnection.

\item[v)]  \inlinecite{Morgan13} have found that the expansion of closed AR loops, discovered by \inlinecite{Uchida92}, continues well above the source surface at 2.5 \Rsun up to 12 \Rsun.  
However, doubts have been raised by \inlinecite{Gopalswamy13} 
about how high-charge states found in the slow wind can be generated by steadily expanding AR loops without requiring magnetic reconnection.  It is also difficult to understand in this scenario why occasionally slow-wind signatures only appear on one side of the heliospheric plasma sheet (HPS) as the expansion of loops would imply symmetrical slow-wind streams on both sides.

\end{enumerate}

A different approach was taken by the high-resolution spectroscopy community.
Since the discovery of persistent hot plasma upflows from the edges of solar
active regions (ARs) by the {\it Hinode}/{\it X-ray Telescope} (XRT:
\opencite{Sakao07}) and the {\it EUV-Imaging Spectrometer} (EIS:
\opencite{Harra08}), {it has been suspected that these upflows are in fact outflows
that contribute to the slow wind.}
\inlinecite{Baker09a} demonstrated that upflows occur at specific
locations of an AR magnetic topology, along QSLs where current
sheets form and magnetic reconnection takes place. A pressure gradient,
resulting from reconnection between loops of different density, was proposed as
a driver of the plasma upflows. Using potential-field extrapolations,
\inlinecite{DelZanna11} linked the upflows of an AR to a high-altitude coronal
null point in a pseudo- (or unipolar) streamer configuration, where
interchange reconnection can take place, creating a large pressure gradient and a
rarefaction wave in the reconnected loops (\opencite{Bradshaw11}). As separatrices and
nulls are part of the more general QSL concept, the results of
\inlinecite{Baker09a} and \inlinecite{DelZanna11} are not contradictory, and they
both highlight the key role of the magnetic-field topology and magnetic
reconnection in the generation of AR plasma upflows. However, the specific
manner in which the plasma upflows of an AR can find their way into the solar
wind was, and still is, a key open issue.

\citeauthor{Brooks11} (\citeyear{Brooks11}, \citeyear{Brooks12})
have determined the FIP bias in AR outflows using {\it Hinode}/EIS data and
found matching values in the slow wind observed with the {\it
Advanced Composition Explorer} (ACE) three days later, which is a promising result. 
Their work supports the existence of a physical link between the plasma outflows in the AR
and the slow-wind stream. Such a connection has been established by
\inlinecite{vanDriel-Gesztelyi12} 
through the use of coronal and \textit{in-situ} observations, combined with
magnetic modelling. They analysed a complex of two active regions flanked by
two low-latitude CHs of opposite magnetic polarity observed between
2\,--\,18 January 2008. Linear force-free (LFF) magnetic-field extrapolations
of the two ARs confirmed that their plasma outflows are co-spatial with the
locations of QSLs, including the separatrix of a magnetic null-point.
Complementarily, a global potential-field source-surface (PFSS) model indicated that 
the smaller AR was only partially covered by a large closed-field streamer. 
At the null point, present above this AR, interchange reconnection  between
closed high-density magnetic loops of the small AR and open evacuated field lines of
the neighbouring CH could take place. The resulting pressure gradient
led to upflows along the reconnected open-field lines, and downflows along
the newly created closed loops, as proposed by \inlinecite{Baker09a} and
modelled by \inlinecite{Bradshaw11}. Using ACE data,
\inlinecite{vanDriel-Gesztelyi12} showed that the temperature
[$T$], velocity [$v$], and ion abundances 
observed in the corresponding slow-wind streams were related to this small parent AR. 

In the example analysed by \inlinecite{vanDriel-Gesztelyi12}, 
the presence of two bipolar ARs creates a quadrupolar magnetic configuration in which a
high-altitude null point could be present. Under favorable circumstances,
the spine of this null and the QSLs that surround it can provide a natural
channel for the AR outflows into the slow wind. However, when 
only one simple bipolar AR is present on the Sun and it
is fully covered by the separatrix of a streamer, as in the configuration studied by
\citeauthor{Brooks11} (\citeyear{Brooks11}, \citeyear{Brooks12}) in December
2007, the question arises: is it possible for the plasma from the AR to
find its way into the slow SW? 
\inlinecite{Culhane14} revisited the results of 
\citeauthor{Brooks11} (\citeyear{Brooks11}, \citeyear{Brooks12}), when the only AR present on the visible disc was 
NOAA AR 10978.
Strong upflows, of up to 50 km s$^{-1}$, were
observed with EIS at both sides of this AR during its disc passage (\opencite{Bryans10};
\opencite{Brooks11}; \opencite{Demoulin13}). 
In spite of the AR being fully covered by a large streamer separatrix, \inlinecite{Culhane14} found signatures typical of AR plasma in the
presumably associated slow-wind stream, which was asymmetric around the HPS. 
How can the coronal plasma circumvent this topological obstacle? In the following sections, we intend to answer this question.  

This article is organised as follows: In \sect{environment}, we briefly describe
the photospheric and coronal environment surrounding AR 10978.
\sect{model-topo} presents a global magnetic-field coronal source-surface model,
which we use as the boundary condition for a detailed computation of the magnetic
topology of the AR and its surroundings. As a result of our topological
analysis, in \sect{escape}, we discuss the way that the AR plasma can escape from
below the closed streamer configuration {\it via} magnetic reconnection
occurring in, at least, two steps. We look for coronal signatures of the two-step reconnection process 
analysing data from the {\it Nan\c cay Radioheliograph} (NRH) (\sect{radio}), which is sensitive to low-energy
release processes. Finally, in \sect{summary}, we discuss the
implications of our topological analysis for the origin of the unsteady slow SW, and we draw some conclusions. 

\begin{figure}  
\centerline{ \includegraphics[width=1.\textwidth]{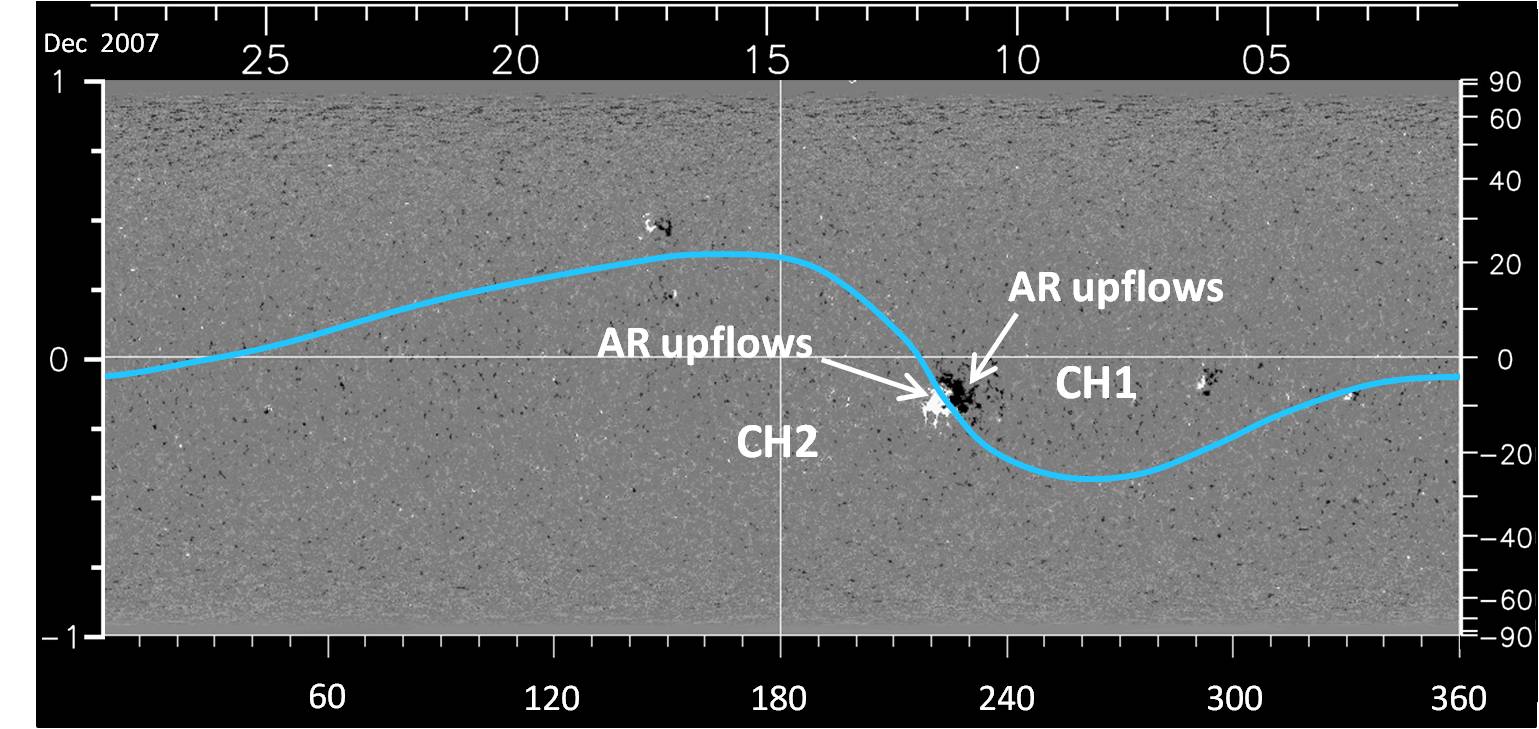} }
\caption{MDI synoptic magnetogram for CR 2064. The horizontal axis
indicates the Carrington longitude, the vertical axis on the right corresponds to
the Carrington latitude, while the one on the left refers to its sine. The thick 
light-blue line overlying the figure is the projection on the disc of the inversion
line at 2.5 \Rsun\ computed {using} the NSO/GONG PFSS model. In passing over the AR,
this projection roughly aligns with the main polarity-inversion line of AR
10978. The CHs at both sides of the AR are indicated, as well as the sites of
most prominent upflows observed by EIS. For all figures showing
magnetic field data, white (black) corresponds to positive (negative) field
values.}
\label{fig:mdi-cmap}
\end{figure}  

\section{The Magnetic and Coronal Environment of AR 10978}
\label{sec:environment}

\subsection{Global Observations During CR 2064}
\label{sec:data}

During Carrington rotation (CR) 2064, AR 10978 was the largest catalogued region present on
the solar disc. This bipolar region transited the disc from 6 to 19 December 2007 as seen from the Earth. It 
had an average total unsigned magnetic flux of $6.0\times$10$^{22}$ Mx. \fig{mdi-cmap} shows a {\it
Michelson Doppler Imager} (MDI: \opencite{Scherrer95}) synoptic map in which AR
10978 is isolated, with only a few other minor regions visible at some distance.  The projection on the solar surface of the
polarity inversion line (PIL) at 2.5 \Rsun\ is shown overlying the synoptic map. 
In passing over AR 10978, the projection of the PIL roughly agrees with the magnetic inversion line of this AR. Two unipolar areas of opposite polarity, corresponding to two CHs to the West and East of the
AR, respectively, are indicated as CH1 and CH2 in \fig{mdi-cmap}. The locations of the peripheral EIS upflows, which are
relevant to our study {(see, {\it e.g.}, \opencite{Demoulin13})}, are also shown. 

\begin{figure}  
\centerline{\includegraphics[width=0.8\textwidth]{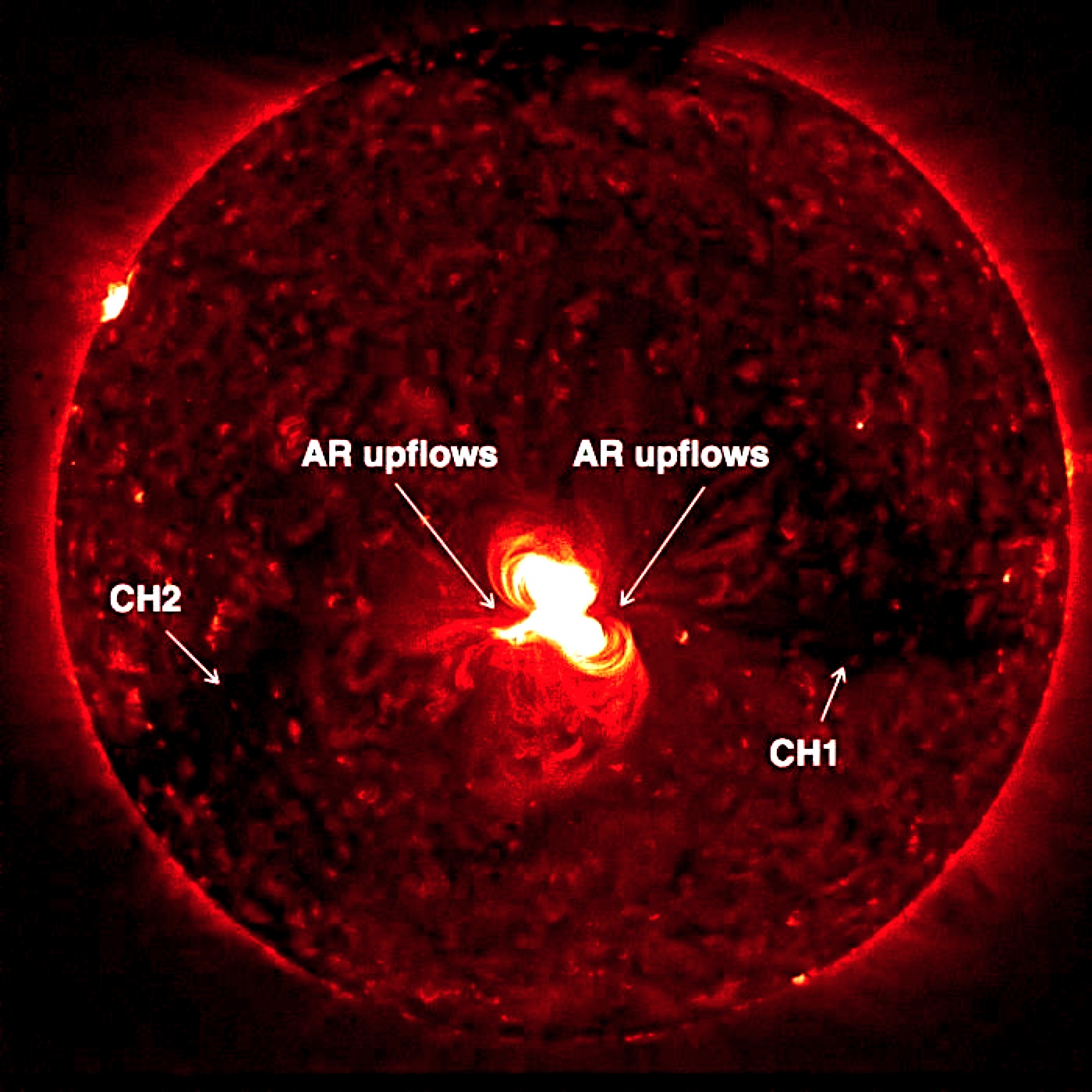}}
\caption{{\it Hinode}/ XRT image showing the {hot} coronal environment around AR 10978.
The image corresponds to the central meridian passage (CMP) of the AR,
on 12 December 2007. Two CHs, indicated by arrows, are seen flanking the region,
which is the only sizable bright feature on the solar disc. The locations of
the main upflows observed by EIS are also indicated by arrows. }
\label{fig:xrt-comp}
\end{figure}

The {hot} coronal environment around AR 10978 is revealed by the X-ray image
shown in \fig{xrt-comp}. The two CHs are clearly seen as
dim regions at both sides of the AR, which is the only significant bright
feature on the disc. 

AR 10978 was numbered as AR 10980 during the next Carrington rotation (CR 2065).
The comparison of the photospheric environment of these two recurrences of the same AR reveals some differences. Although a few small bipoles emerged during 
the disc 
transit of AR 10978, they did not alter its strongly bipolar configuration. On
the other hand, AR 10980, while being bipolar and already in its decay phase,
was closely followed by a spotless bipolar region that appeared at the eastern
limb on 3 January 2008 (\opencite{vanDriel-Gesztelyi12}). The two regions coupled forming a quadrupolar magnetic configuration. {Both AR 10978 (CR 2064) and AR 10980 and its coupled small region (CR 2065), were flanked by 
CHs and remained stable during their disc transits. A series of 
C-class X-ray flares occurred in AR 10978, at a rate of two {\it per} day, after 13 December
and until its disappearance behind the western limb. 
These flaring episodes are related to the emergence of small weak bipoles at both
sides of the main negative polarity (see Figure 1 in \opencite{Demoulin13}).
Only one C-class flare occurred in AR 10980 on 7 January 2008 with an associated small-scale 
coronal mass ejection (CME) (see \opencite{Foullon11}).}

\subsection{The Global Field Model of CR 2064} 
\label{sec:FDIPS}

\begin{figure}  
\centerline{\includegraphics[width=0.9\textwidth]{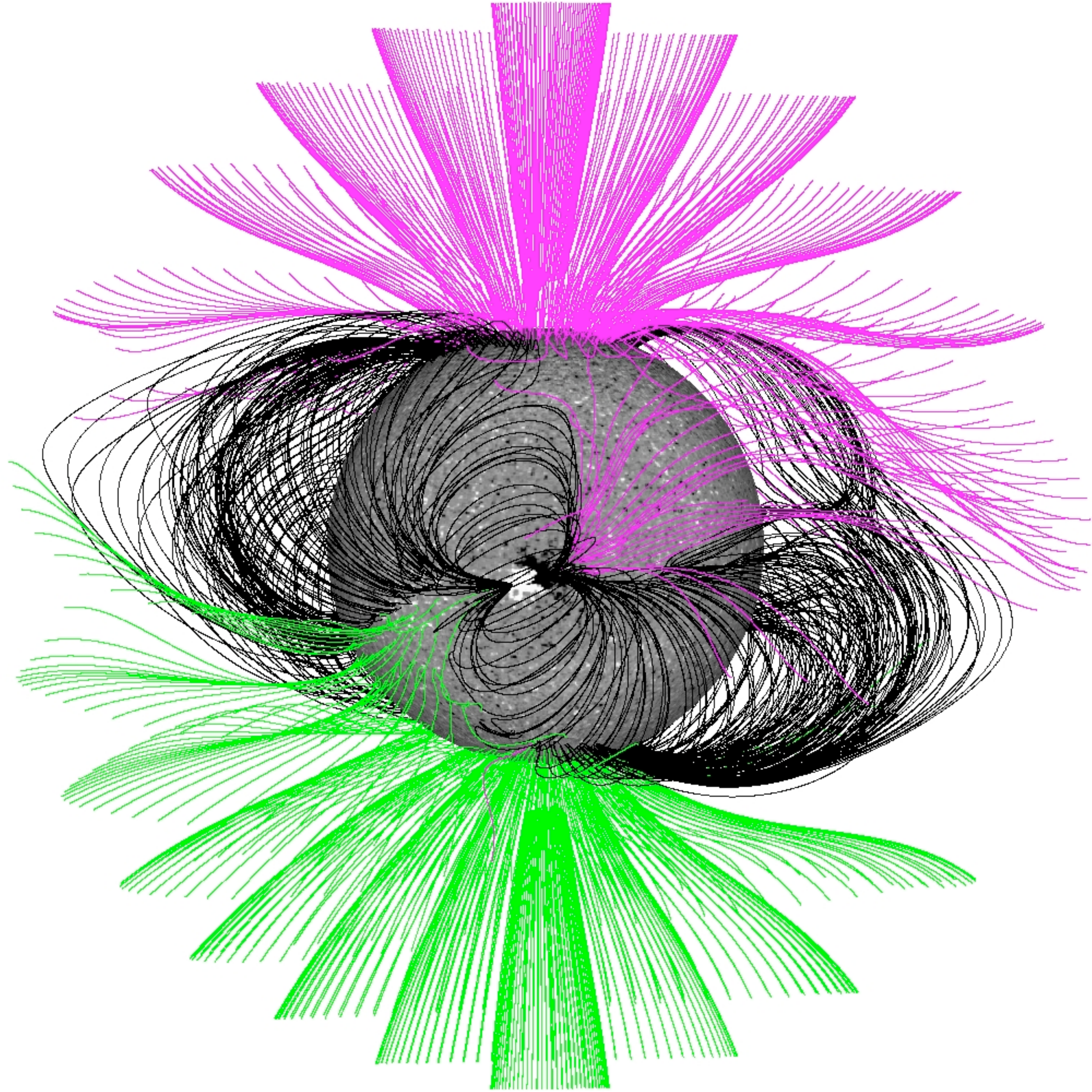}}
\caption{PFSS model of CR 2064 with AR 10978 at CMP (Carrington longitude
$228\degree$). The AR lies completely below the streamer belt and open-field lines of the two CHs are seen at both of its flanks. The field-line colour convention is such that:
black indicates closed lines and pink (green) corresponds to open lines anchored in the
negative-polarity (positive-polarity) field.}
\label{fig:global}
\end{figure}

The global coronal magnetic field of CR 2064 is modelled using a PFSS approach.
These models assume a current-free coronal field with an observationally
prescribed and/or modelled boundary condition at the photosphere. PFSS models
assume that the field becomes purely radial at a given height called the
source surface: a free parameter usually set to the value 2.5 \Rsun. The PFSS
model used in this article (see \fig{global}) was computed with the Finite
Difference Iterative Potential-Field Solver (FDIPS) code described by
\inlinecite{Toth11}, using the MDI synoptic magnetogram for CR 2064
as photospheric boundary condition. 

Individual full-disc MDI magnetograms are built every 96 minutes. The
calibrated photospheric line-of-sight (LOS) field is transformed into the total field strength, assuming the magnetic-field vector is fully radial at the location of the measurement. To construct a synoptic map of the entire solar
surface, the field strength is averaged at each grid point from the time series of $\approx 20$ magnetograms, which provide data for that grid point in a $\pm(\approx10^\circ)$ longitudinal range around its CMP. Prior to averaging, the individual magnetograms are interpolated to disc-centre resolution, resulting in a $3600\times1080$ pixel synoptic map, and differential rotation is accounted for. The axes are linear in Carrington longitude (0.1-degree intervals) and in sine latitude, so that each synoptic pixel represents
the same area on the solar surface. The LOS projection effect severely
compromises the measured magnetic data at large latitudes. In the latitude range
$\pm 60^\circ$, the synoptic maps preserve only observed data. For higher
latitudes, special sets of observations and interpolation schemes are used. Full
details concerning the construction of synoptic maps can be found in the MDI web-site (\url{http://soi.stanford.edu/magnetic/index.html}). It is important to keep in mind that synoptic magnetograms are an average
description of the magnetic-flux distribution at the solar surface; therefore, they
provide a meaningful description for regions that show stability over the
averaging period, which is the case for AR 10978.

PFSS models are traditionally computed using a spherical-harmonic
decomposition of synoptic magnetograms. These models, if not properly apolized,  are characterised by
high-frequency artefacts around sharp features.
They can also lead to inaccurate and unreliable results when using a grid that
is uniform in the sine of the latitude, especially in the polar regions close to
the Sun's surface. Alternatively, the FDIPS code, which is freely available
from the Center for Space Environment Modeling (CSEM) at the
University of Michigan (\url{http://csem.engin.umich.edu/tools/FDIPS}),
makes use of an iterative finite-difference method to solve the Laplace equation
for the magnetic field. While the spherical harmonics are global functions and
their amplitudes depend on all of the magnetogram data, these data affect the
solution only locally in the finite-difference approach, which leads to a
solution that behaves better in the presence of large spatial gradients. As a
result, the finite-difference method provides more accurate results at solar
latitudes higher than those typically used in the spherical harmonic expansion
methods (\opencite{Toth11}).

The FDIPS model runs on a spherical grid that is uniform in longitude, sine
of latitude, and radial direction. For the present work we ran the model
with a resolution of $0.5^{\circ}$ in longitude (720 longitudinal grid points),
$9.3\times 10^{-3}$ in the sine of latitude (216 latitudinal grid points,
which implies an effective latitude resolution of $0.5^\circ$ to $1.0^\circ$ in the
$\approx\pm 65^\circ$ latitude range), and $5\times 10^{-3}$\Rsun\ in the radial
direction. Note that the dimension in each angular direction has been chosen to
be an integer fraction of the respective dimension in the MDI magnetogram. In
this way, the cells of the magnetogram can be merged together without losing
magnetic flux.

The coronal magnetic field is globally well described by PFSS models, except in
coronal regions where high currents are present, such as could be the case around filaments. 
However, for the purposes of our work, a PFSS model provides a reasonable
description of the connectivity of the magnetic field around the isolated and
stable AR 10978. This is illustrated in \fig{global}, which shows the result of
our PFSS model when AR 10978 was at the central meridian.
The displayed field lines have been traced from a set of coronal starting
points  separated by $10^\circ$ (in both latitude and longitude) and are located
at a height of $\approx 50$ Mm above the photosphere. This selection provides
a fairly uniform sampling of the large scale structures, especially of the streamer
arches above the AR. The magnetic field line passing each selected starting
point is then traced in both directions until the {solar atmosphere base} (1.0 \Rsun) or
the source surface (2.5 \Rsun) is reached. The tracing is performed
using a forward Euler method software
package (\url{http://www.lmsal.com/~derosa/pfsspack}, 
spherical\_trace\_field.pro). 
Two points are noteworthy in \fig{global}. First, a
good agreement between the locations of the CHs and the AR loops is evident when
comparing \fig{global} with \fig{xrt-comp} and, second, AR 10978 is fully covered by the
streamer-belt arcade {(see also Figures 5a, 6a, and 6b in \opencite{Culhane14}). The latter 
suggests that it is very difficult for the 
plasma in EIS upflows, {which cannot access open-field lines}, to find its way out 
into the slow SW despite the findings of \inlinecite{Culhane14}.}

\section{Global Magnetic Field Topology of CR 2064}
\label{sec:model-topo}

\subsection{Coronal Magnetic Null-Points}
\label{sec:nulls-comp}

Since the seminal works on the origin of flares (see, {\it e.g.}, 
\opencite{Sweet58}), it has been thought that null points are the 
topological structures in the solar corona in 
which magnetic-field reconnection and energy release is most likely to occur.
In view of the results of \inlinecite{vanDriel-Gesztelyi12}, who found that
magnetic-field lines in the vicinity of a null point can undergo interchange reconnection 
and channel AR plasma into the slow SW, we search for the presence of null
points in CR 2064. 

The field connectivity in the neighbourhood of a null point displays a structure characterised by the so-called spines and fans (see, {\it e.g.}, 
\opencite{Longcope05b};
\opencite{Pontin11}). Several works have analysed, either numerically
(\opencite{Craig96}, \citeyear{Craig99}; \opencite{Wyper13}) or analytically
(\opencite{Ji01}; \opencite{WilmotSmith11}), solutions for fan and spine magnetic 
reconnection. Observationally, the origin of
several flares can be attributed to magnetic reconnection in the fan or spine
structure of null points (\eg\ \opencite{Mandrini91}, \citeyear{Mandrini93}, \citeyear{Mandrini06}, \citeyear{Mandrini14};
\opencite{Parnell94}; \opencite{Aulanier00}; \opencite{Manoharan05};
\opencite{Luoni07}; \opencite{Reid12}); although others are not related at all to their presence
(\eg\ \opencite{Demoulin94b}; \opencite{Mandrini96}; \opencite{Bagala00};
\opencite{Schmieder07}; \opencite{Savcheva12}).      

{From a mathematical point of view, the neighbourhood of a magnetic
null point can be described by the linear term in the local Taylor
expansion of the magnetic field (see \opencite{Demoulin94b}, and references
therein). Diagonalisation of the Jacobian field-matrix gives three
eigenvectors and the corresponding eigenvalues, which add up to zero in
order to locally satisfy the divergence-free condition on the field. 
Under coronal conditions, the eigenvalues are real (\opencite{Lau90}).
A positive null point has two positive fan eigenvalues 
and conversely for a negative null.
When a null point is present the coronal volume is divided into two
connectivity domains, which are separated by the surface of the fan. In each of these
domains a spine is present. 
We will refer to the lower (upper) spine as the one arising
from below (above) the fan surface. This surface is defined by all of the field lines that start at an infinitesimal
distance from the null in the plane defined by the two eigenvectors associated
with the eigenvalues having the same sign.} 

\begin{figure}  
\centerline{\includegraphics[width=1.\textwidth]{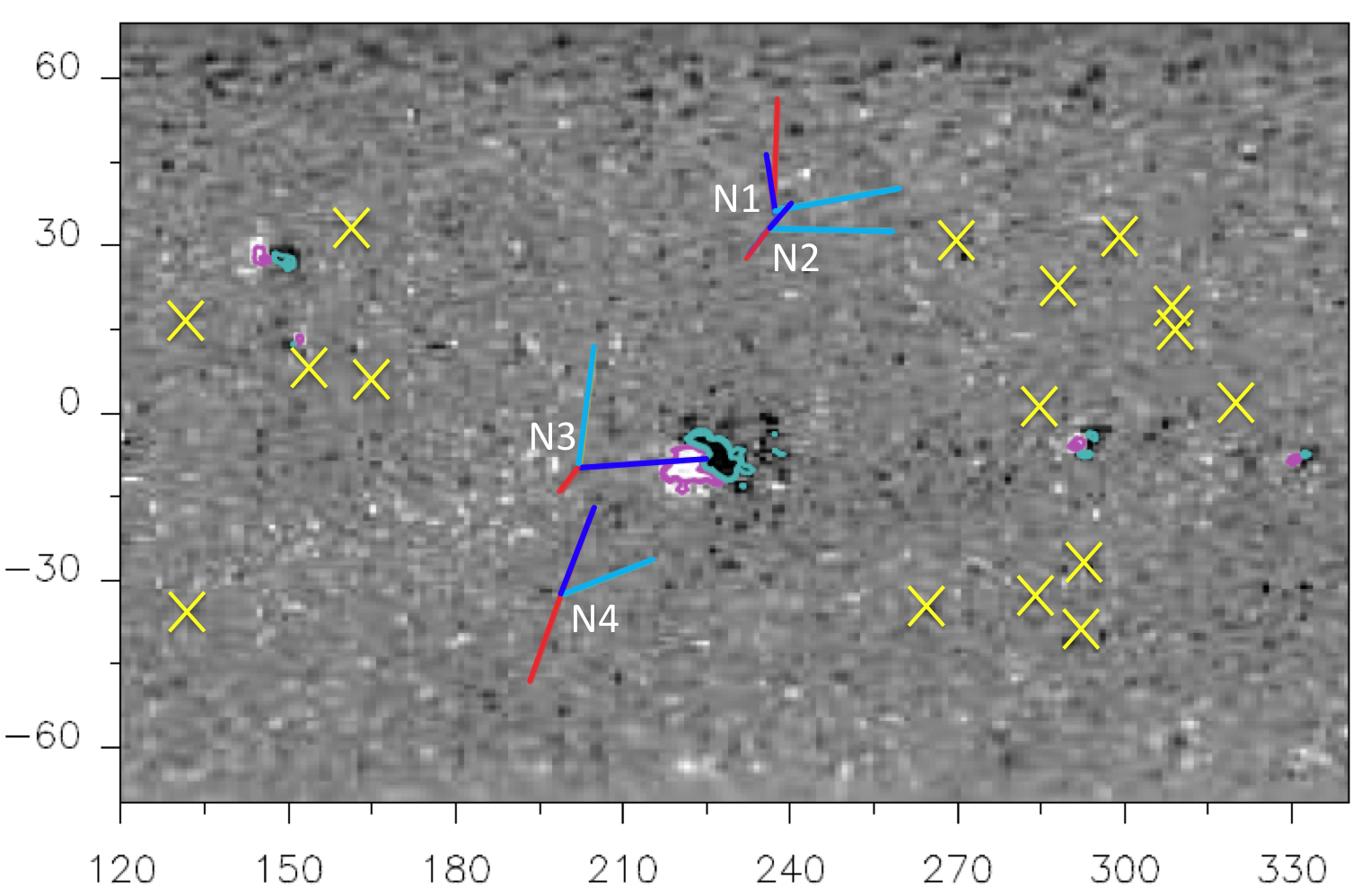}}
\caption{
A section of the synoptic map of CR 2064. The magnetic field, which
corresponds to the photospheric boundary of the PFSS model, is saturated above
(below) 30 G (-30 G). The locations of all magnetic null-points at
heights greater than 0.036 \Rsun\ ($\approx$ 25 Mm) above the photosphere are
shown either as the intersection of a set of three coloured segments (see text) when they are 
within $20\degree$ in longitude from AR 10978, or with a yellow $\times$-symbol when they are at greater longitudes.  
Nulls identified by the three segments are also labeled with a letter and a number. The horizontal
(vertical) axis is the Carrington longitude (latitude).}
\label{fig:null-loc}
\end{figure}

The method to locate the magnetic null-points in CR 2064 is similar to that
discussed by \inlinecite{Demoulin94b} but using a spherical geometry for the
coronal-field model. For a large sub-area of the whole photospheric map shown in
\fig{mdi-cmap}, \fig{null-loc} shows the location of all of the null points found
at a height of 25 Mm or greater above the photosphere within the displayed range. 
The threshold height of 25 Mm has been arbitrarily chosen for the figure. We have carried out
several searches for nulls using different threshold heights and found that
their number decreases rapidly with the height above the photosphere, as
was shown by \inlinecite{Longcope09} for PFSS models constrained by
quiet-Sun MDI magnetograms, and \inlinecite{Schrijver02} using a modeled
boundary condition consisting of a random distribution of magnetic sources. 
Furthermore, for any threshold height $\lesssim$ 25 Mm
most of the null points found are not related to AR 10978 or to the EIS
upflows at its edges, or they present a closed-magnetic field structure (as we
will show in \sect{nulls-closed} for some of the nulls in \fig{null-loc}).
These characteristics do not make those nulls good candidates for magnetic
reconnection enabling the plasma of EIS upflows to reach the SW; therefore,
they can be ignored.

\begin{figure} 
\centerline{\includegraphics[width=0.8\textwidth]{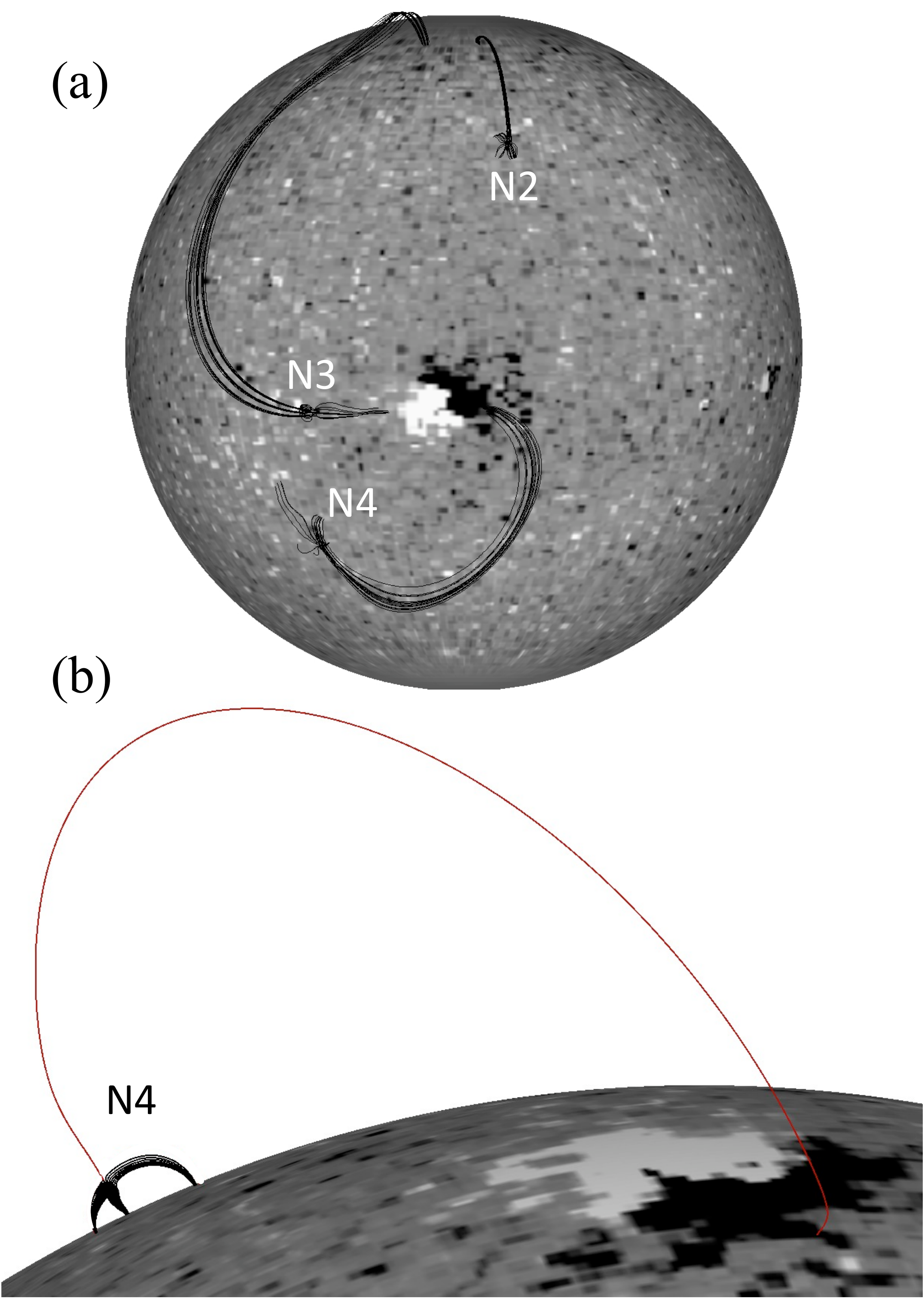}} 
\caption{Magnetic nulls with a large-scale and closed magnetic configuration
around AR 10978.
(a) Nulls 2, 3, and 4 and a set of field lines computed from their surroundings,
as explained in \sects{nulls-comp}{nulls-closed}.  The field lines drawn around
N3 are connected to the network close to AR 10978, while those around N4 reach
its western negative polarity. All field lines (drawn in black colour) computed
for this set of nulls are closed. 
(b) A close-up and rotated view of null N4 illustrating a set of fan field lines and the 
upper spine line drawn in black
and in red colours, respectively.}
\label{fig:closed-nulls}
\end{figure}  

Following \inlinecite{Demoulin94b} (see also \citeauthor{Mandrini06}
\citeyear{Mandrini06}, \citeyear{Mandrini14}), \fig{null-loc} shows the location of
all null points within an angular distance of $20\degree$ in longitude from AR 10978 as the intersection of a set of three coloured segments, which correspond to the
direction of the three eigenvectors of the Jacobian matrix. 
The colours of these segments indicate the magnitude of the corresponding
eigenvalue: Red (yellow) corresponds to the largest (lowest) positive eigenvalue
in the fan plane and blue to the spine eigenvalue for a positive null point. For
a negative null, dark blue (light blue) corresponds to the largest (lowest)
negative eigenvalue in the fan plane and red to the spine eigenvalue. These null points have been also
identified with a letter and a number, and  their magnetic structure will be analysed in
detail in \sects{nulls-closed}{nulls-open}.
The locations of nulls farther than  $20\degree$ in longitude from AR 10978 are indicated with a yellow $\times$-symbol.

In order to test the stability of the nulls identified in \fig{null-loc}, we
have computed their locations, eigenvectors, and eigenvalues using two
alternative PFSS models: i) doubling the resolution of the previous model, and
ii) keeping the same resolution but changing the boundary condition to a 
synoptic map of CR 2064 from the Global Oscillation Network Group (GONG). GONG 
maps have a lower resolution, about ten times
coarser than that of MDI, with 360 data points linear in longitude (or a
resolution of $1^{\circ}$) and 180 data points linear in sine of latitude
(against the 1080 data points of MDI). In performing experiment i), we found
that the change in resolution kept the location of the nulls unaltered within a
precision of $\approx10^{-3}$ and preserved the characteristics of their
topology (signs and relative values of the eigenvalues). 
In performing experiment ii), we found a similar result, with the location of the
nulls being unaltered in radius and latitude within a precision of
$\approx10^{-2}$, and in longitude within a precision of $\approx10^{-1}$. From
these results, we conclude that the null points are indeed stable and their presence and characteristics are not an artefact of the particular data set used as the boundary
condition for the model and/or its numerical resolution.

\subsection{Nulls with Closed Magnetic Configuration in AR 10978 Neighbourhood}
\label{sec:nulls-closed}

\begin{figure} 
\centerline{\includegraphics[width=\textwidth]{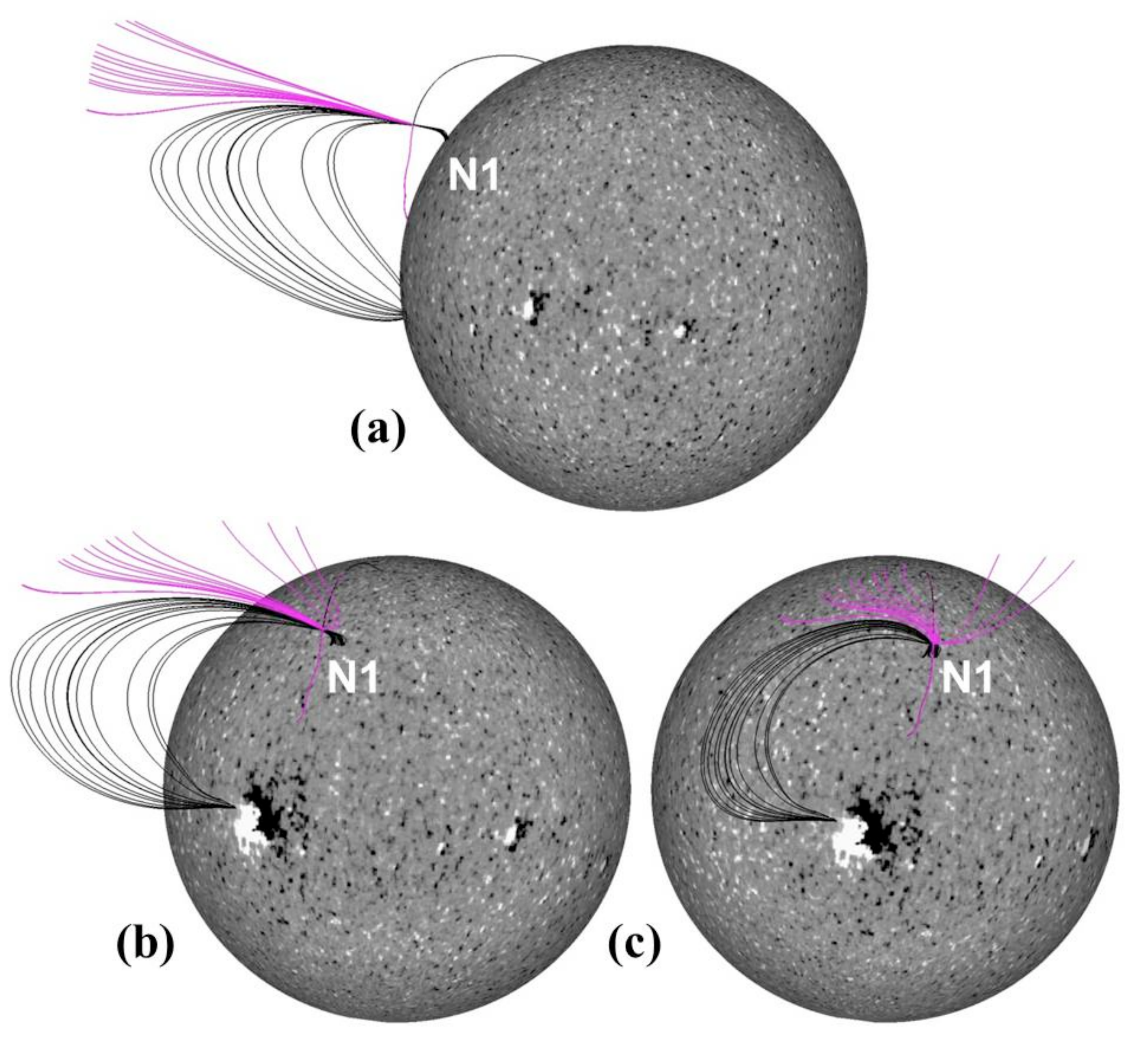}} 
\caption{
Different views of a set of field lines that pass in the neighbourhood of
null point N1. The panels are different views of the same model. From the set of null
points selected within $\pm 20\degree$ in longitude from AR 10978, null N1 is
the only one associated with an open magnetic-field configuration. (a)
Corresponds to what an observer facing the Sun would see when the center of AR 10978 was at the East limb.  (b) and (c) correspond to what the same observer would see when the center of the AR was at a solar longitude of  
$45^\circ$ and $0^\circ$ , respectively.  Open field
lines anchored in negative-polarity field are drawn in pink, while closed-field
lines are drawn in black.  Note that in (a) the open-field lines, rooted in the negative polarity, shift towards the Ecliptic at the source surface as their photospheric footpoints shift slightly towards the solar Equator (compare panel a to b).}
\label{fig:open-null}
\end{figure}

\begin{figure} 
\centerline{\includegraphics[width=0.6\textwidth]{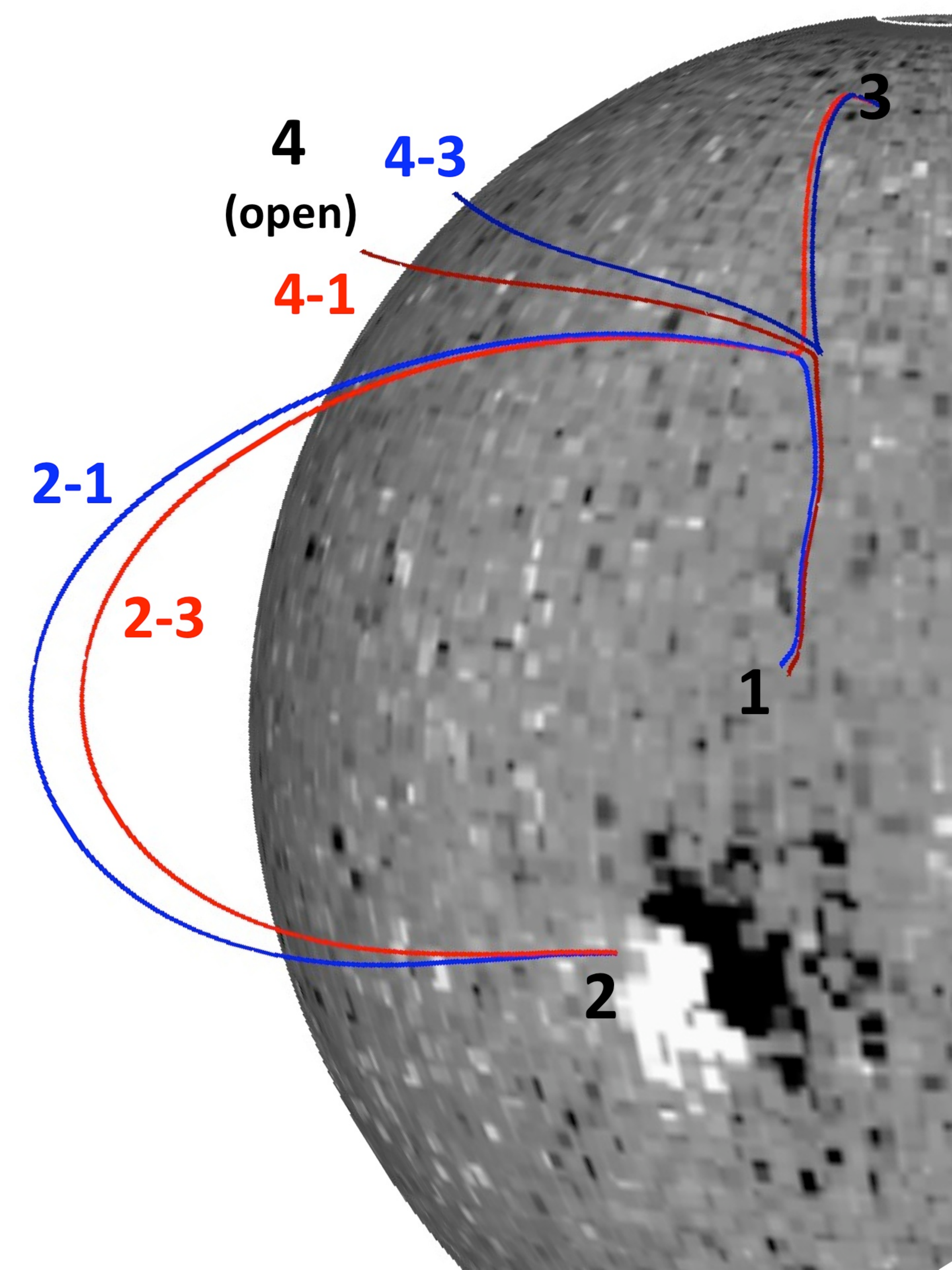}} 
\caption{A selection of four field lines computed from the vicinity of null point N1.  These
have been chosen from the set in \fig{open-null} to illustrate how
interchange reconnection could proceed at the null. Even (Odd)
numbers at the photospheric level indicate positive (negative) polarity regions
to which the field lines are connected. Number 4 is used for the source surface
at 2.5 \Rsun . Numbers separated with a dash,
set by each field line, indicate its connectivity. The field lines have been
coloured such that: 
blue corresponds to pre-reconnection closed and open field lines,  
while red corresponds to post-reconnection closed and open-field lines, respectively.
}
\label{fig:null1-recon}
\end{figure}

In Figures \ref{fig:closed-nulls} to \ref{fig:null1-recon}, we
show selected field lines that pass through locations in the vicinity of each
null, N1 to N4. 
At each null, the eigenvectors that correspond to the spine line and the fan
plane are computed. This information is used to select a number of position
vectors around each null point, to be used as starting points to trace field
lines.

\fig{closed-nulls}a shows the magnetic connectivity in the vicinity 
of nulls N2, N3, and N4, which are all negative null points, located above the
photosphere at heights 25.2, 26.4, and 25.7 Mm, respectively. To trace field
lines for this figure we selected several starting points around each null,
above and below their fan planes. 
The selected field lines clearly show that their large-scale magnetic
configuration is closed.  The upper spine of null N2 connects far away, towards the
northern CH. Field lines around the upper spine of null N4 connect to the
western negative polarity of AR 10978, while some field lines around N3 reach the
network close to its eastern positive polarity at one footpoint and the northern
CH at the other footpoint.

\fig{closed-nulls}b shows, as an example, a set of the fan field lines of null
N4. To trace field lines, we selected several starting points in the fan plane, which were computed as the null position plus linear combinations of the two fan eigenvectors. We also selected a starting point along the spine line, computed as the null position plus an infinitesimal departure along the spine eigenvector. 
The field lines chosen to start in the fan plane are shown in black, while
the upper spine line is shown in red.

\subsection{Nulls with Open Magnetic Configuration in the Neighbourhood of AR 10978}
\label{sec:nulls-open}

The only magnetic null point associated with an open magnetic-field
configuration is null N1 (\fig{open-null}).  It is located to the North-West of the
main negative polarity of AR 10978, close to the northern negative CH.  Its
height above the photosphere is $\approx$ 108 Mm. The starting points to trace
the field lines around this null have been selected on a $3\times7\times7$
(radial$\,\times\,$latitudinal$\,\times\,$longitudinal) spherical grid centered at its
location, with a 0.02 \Rsun\ grid step in the radial direction and a $0.5^\circ$
step in both angular directions. Only a subset of 49 points with heights greater 
than that of the null are chosen for the field lines drawn in \fig{open-null},
where they are shown from three different points of view. 
\fig{open-null}a shows that the open-field lines around the upper spine
slightly shift towards the Ecliptic at the source surface as their footpoints lie closer to the solar
Equator (compare panel a to panel b). The other two points of view correspond to
an intermediate one (panel b) and to the AR seen at its CMP (panel c).  
      
\fig{null1-recon} shows a selection of four field lines from the set in \fig{open-null} in the neighbourhood of
null N1. These selected field lines illustrate
how reconnection could proceed at this null. Large-scale loops anchored close to
the following polarity of AR 10978, labelled as 2--1, could reconnect with 
open-field lines from the northern CH, labelled as 4--3, to give ({\it via}
interchange reconnection) field lines labelled as 2--3  and 4--1. This process
would transfer the closed-loop coronal plasma to the open-field and displace the footpoints of the open-field lines towards the solar Equator.  As a consequence, at the source surface these reconnected open-field lines would shift towards the Ecliptic.  However, this one-step reconnection process is not sufficient to 
bring the plasma contained in the AR upflows into the open field 
{(see \sect{escape})}.

\begin{figure} 
\centerline{\includegraphics[width=0.8\textwidth]{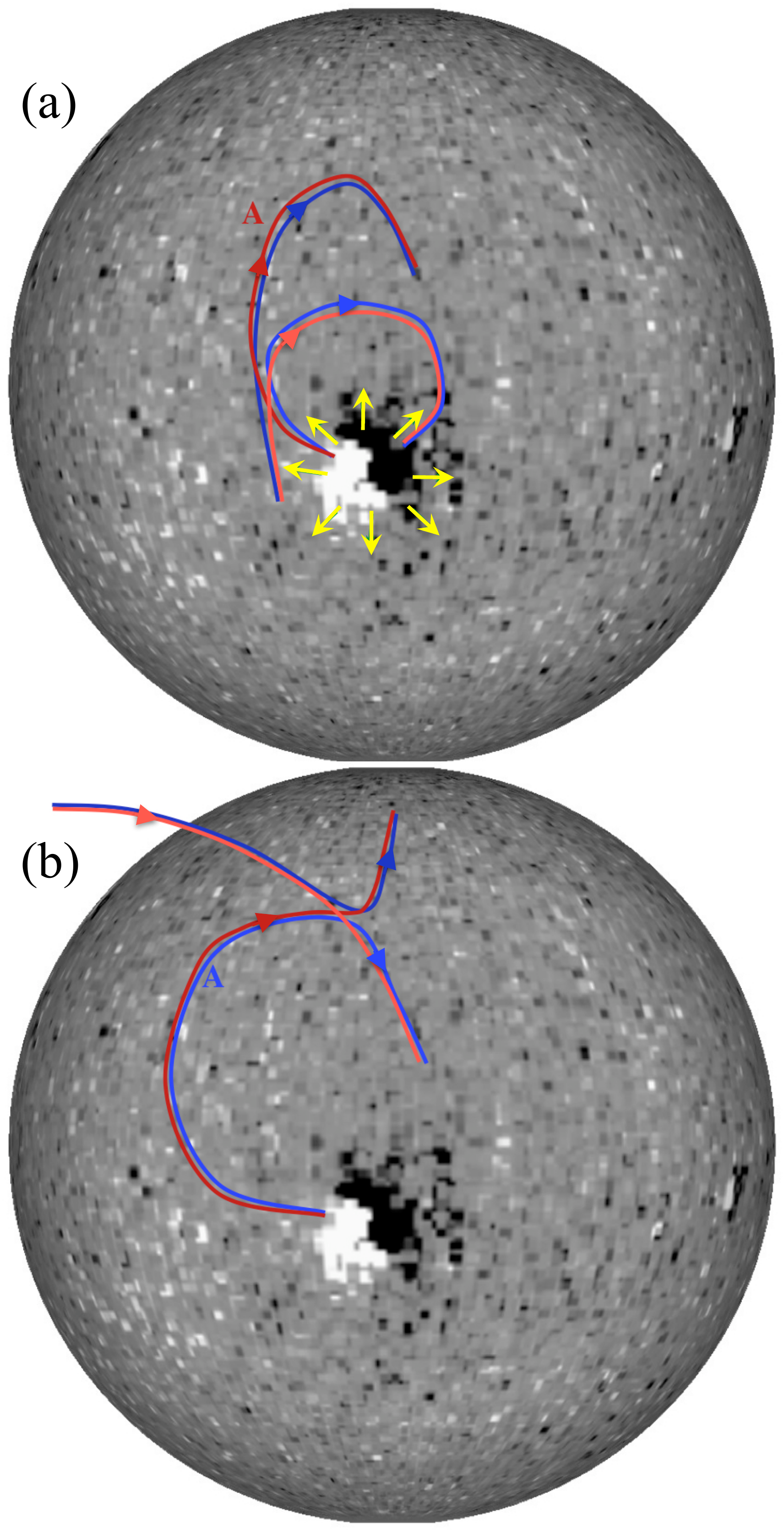}} 
\caption{Sketch showing how the AR plasma can be injected into the solar wind
after a minimum of two main reconnection steps.  
In blue continuous lines we depict the field lines before reconnection, while in red and pink we show the post-reconnection field lines.  
  (a) The AR expansion, indicated by yellow arrows, forces reconnection between the AR closed loops and the network large-scale field lines.
  (b) The further diffusion of the photospheric AR field induces reconnection with the open field of the northern CH. As a result, plasma of AR origin injected in the large-scale closed loops by the first reconnection, can reach the solar wind.   
The reconnected open field lines progressively bend towards the Ecliptic (see \fig{open-null}) and the AR plasma can be later detected at 1 AU by the ACE spacecraft.}
\label{fig:sketch}
\end{figure}  

\subsection{How Can the AR Plasma Escape into the SW?}
\label{sec:escape}

As discussed in \sect{environment} and illustrated in \figs{xrt-comp}{global}, AR 10978 lies completely below the streamer belt. This means that, even if some field lines anchored close to its following positive polarity could reconnect with the northern CH open lines at null N1 (see \sect{nulls-open}), this interchange reconnection process cannot bring the plasma in the closed loops at the periphery of the AR directly into the SW. 
 
However, the combination of our magnetic topology analysis, discussed in \sect{nulls-open}, with the results of \inlinecite{Baker09a} for a different AR suggests the way that the plasma in EIS upflows can reach the SW. In a first step, the AR expansion {could force} reconnection at QSLs between the AR closed loops at its borders and large-scale network fields anchored to the East of the region; these pre-reconnected field lines are drawn in blue {in the sketch of} \fig{sketch}a. As a result of this process, the reconnected field lines, drawn in red, will connect the positive network field to the East of the AR to its preceding negative polarity and to the network at the North of the region. In this way, the plasma in the EIS upflows, mainly from the eastern side of the AR and to a lesser degree from the western side, {could gain} access to 
field lines similar to that labeled as A (see \fig{sketch}a). These field lines 
represent pre-reconnection field lines in a second reconnection process.    

In a second step, illustrated {in the sketch of} \fig{sketch}b, the further diffusion of the photospheric AR field {could induce} the reconnection of the previous large-scale field lines, in particular field lines similar to A (compare both panels in \fig{sketch}), with the open ones anchored in the neighbourhood of the northern negative CH. This reconnection process {may occur} at null N1, located to the North-West of the AR, and associated separatrices, as we have discussed in \sect{nulls-open}. As reconnection proceeds, the reconnected open field lines {would bend} progressively towards the Ecliptic (see \fig{open-null}) so that plasma of AR origin {could be} later detected at 1 AU. 
{We speculate that the two-step reconnection scenario, based on the finding of a null point associated with field lines that are open toward the interplanetary space in our PFSS model, is the only way through which plasma contained in originally closed AR 10978 field lines can be released into the SW. In practice, more reconnection steps are likely to happen, mainly, within QSLs where a continuous slipping reconnection process has been found to be at work (\opencite{Aulanier06}).}

\begin{figure} 
\centerline{\includegraphics[width=\textwidth]{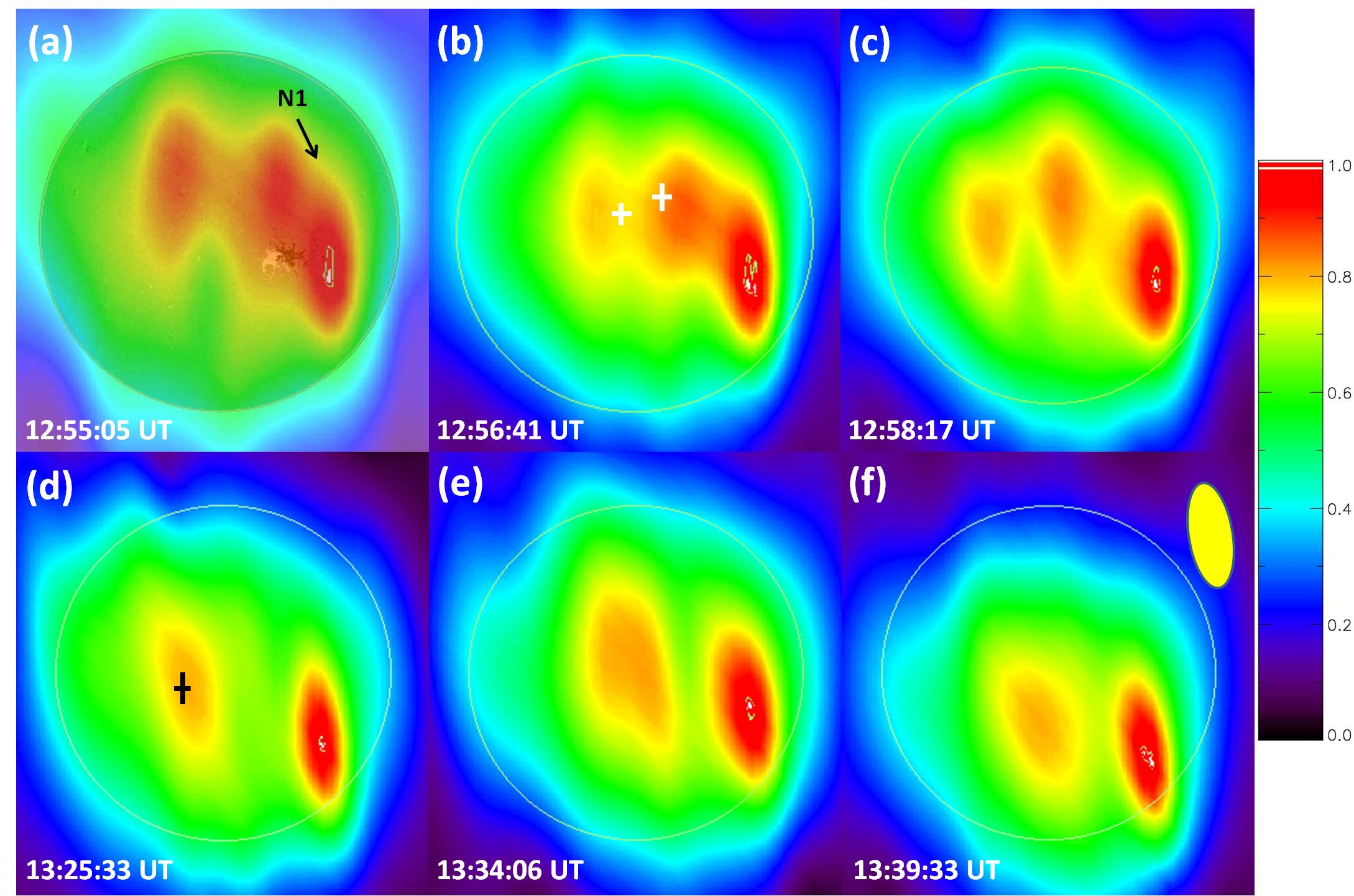}}
\caption{(a), (b), and (c): Brightness distribution over the solar disc at 150.9~MHz in the period 12:55\,--\,13:10 UT
on 13 December 2007. In addition to the main emission associated with the AR (westernmost feature in each image), we observe different locations of dynamically varying radio-emission which eventually reach levels similar to those of the noise storm above the AR.
The white crosses in (b) indicate the average position of the two main fluctuating sources and the uncertainty of their locations is given by the size of the crosses. The image (partially transparent) in panel (a) has been superimposed on the MDI full-disc magnetogram on 13 December 2007 at 12:45 UT. {The arrow points to the location of null N1.} (d), (e), and (f): Similar to the top panels but for the 13:25\,--\,13:40~UT period. The average position of the sources is now indicated by a black cross and its size is a measure of their location uncertainty. {The yellow ellipse in panel f indicates the beam size and its orientation computated at 13:00 UT.} This lobe is computed at half the maximum beam power and its size implies that the observed sources are not resolved. The brightness distribution of each image has been normalized to its maximum value, and the bar to the right indicates the colour scale.}
\label{fig:nrh1}
\end{figure}  

\begin{figure} 
\centerline{\includegraphics[width=0.9\textwidth]{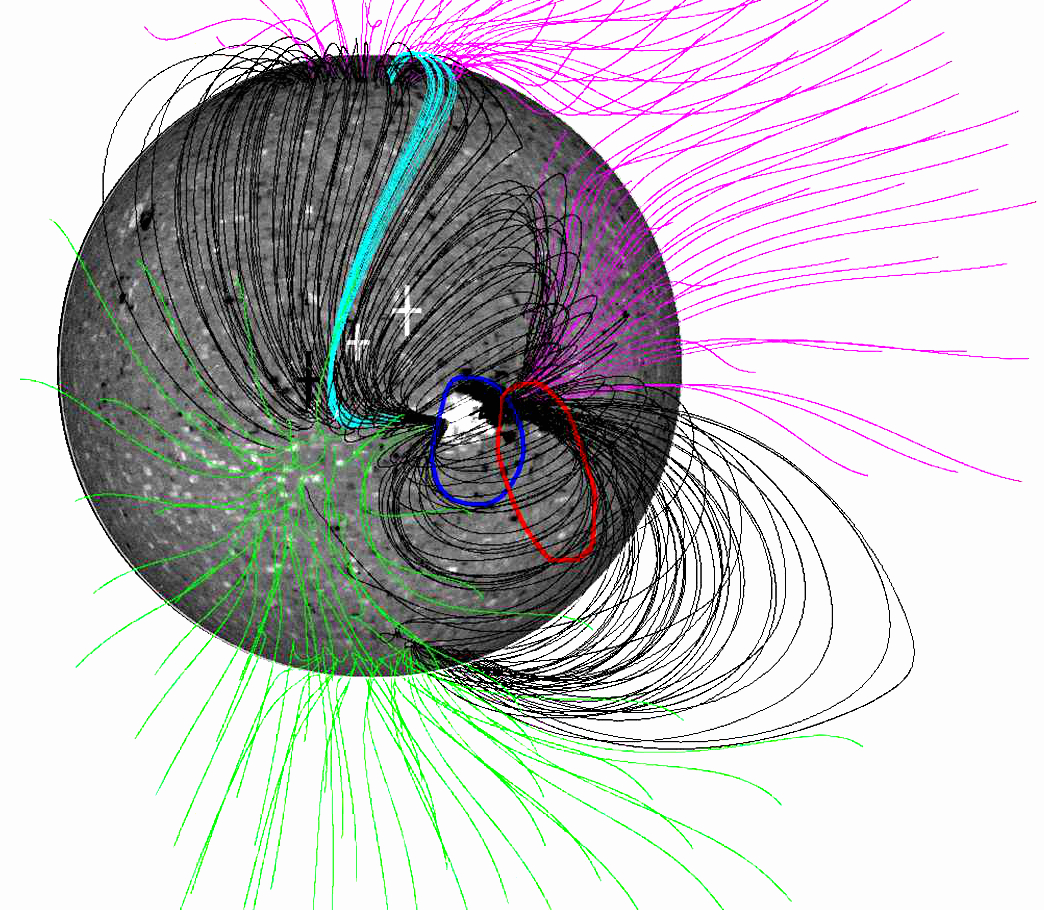}} 
\caption{Magnetic field-lines calculated from the PFSS model rotated to the same time as that of NRH emission isocontours (13:25 UT on 13 December). To the randomly computed field lines starting integration at 50 Mm above the photosphere, we have added a set of lines in the neighbourhood of null point N3, drawn in light-blue colour. The convention for the field line colour, except for those associated with null N3, is the same as in \fig{global}.  
The magnetic field has been saturated above (below) 300 G (-300 G). Superimposed we show isocontours of NRH radio emission at 432 (blue) and 150.9 (red) MHz. Their intensity corresponds to 80\% of the maximum and they have been obtained with $ \approx$ one-minute integration. At the highest frequency, the emission is concentrated over 
the AR and encloses it, while the 150.9 MHz contour appears shifted to the South--West of the AR (see \fig{nrh1}). We have also added the white and black crosses shown in \fig{nrh1}b and d, which are at the average locations of the sporadic sources during the two time intervals studied in detail (see \sect{observations}).}
\label{fig:nrh2}
\end{figure}  

\section{Evidence of the Two-step Magnetic Reconnection Process}
\label{sec:radio}

\subsection{Characteristics of the Radio Emission}
\label{sec:radio-general}

The reconnection processes, described above, occur in regions where the 
magne\-tic-field intensity is low; hence,  we do not expect to find evidence of the energy released in wavelength ranges such as the EUV or the visible. We therefore analysed the radio emission at frequencies sensitive to weak energy release.
Excluding more eruptive transient phenomena, noise storms at metric and 
decimetric wavelengths are the oldest evidence of electron acceleration in 
the solar corona. Radio emission of a noise storm consists of a broad-band continuum lasting from a 
few tens of minutes to a few days with superimposed Type I bursts of short duration and of narrow spectral 
bandwidth (typically a few MHz). They are mainly observed in the 100\,--\,500 MHz frequency range. 
The consensus is that noise-storm emission is triggered by supra-thermal electrons (few keVs), which produce Langmuir waves which are subsequently converted to electromagnetic radiation near the local plasma frequency. This explains the strong polarisation observed in the ordinary mode and also the high value of the 
brightness temperature (see for example \opencite{Kundu:1965}; \opencite{Melrose:1980}; \opencite{Wentzel:1986}). 
Le Squeren (\citeyear{Lesqueren:1963}) determined that the emission at 169~MHz during noise storms {comes from layers located between 60 and 600 Mm (0.09\,--\,0.85 \Rsun) above the photosphere} and, in general, from much higher altitudes 
than those estimated for an unperturbed normal corona. At the same frequency, Kerdraon and 
Mercier (\citeyear{KerdraonandMercier:1982}) found a mean altitude of 130~Mm (0.16 \Rsun) for 
a series of selected noise storms that they analysed. Interestingly, centres of noise storms 
are not situated radially above sunspot groups.

{The {\it Nan\c{c}ay Radioheliograph} (NRH: \opencite{Kerdraon:1997}), with an operational frequency 
range between 150 and 432~MHz, allows imaging of weak energy release at coronal heights such as those 
where noise storms originate. Considering Coulomb collisions with ambient electrons as the energy-loss mechanism, 
\inlinecite{RaulinandKlein:1994} showed that electrons with energies required to produce noise storms have lifetimes of tens of seconds for ambient densities between $3\times10^8$~cm$^{-3}$ and $2\times10^9$~cm$^{-3}$, which correspond to the plasma frequencies in the NRH operational range. These densities are
typical of the low and middle corona ($\approx$~0.1\,--\,0.5 \Rsun~above the photosphere).

Ionospheric effects scale as $\nu^{-2}$ (\opencite{Bougeret:1973}) and are 
mainly due to electron-density fluctuations  
caused by atmospheric gravity waves, which produce travelling ionospheric 
disturbances (TIDs). At metric wavelengths the only effect of TIDs is a periodic shift of the 
radio source location (\opencite{Mercieretal:1989}), which is evident in radio observations made, in general, 
before Noon  (\opencite{SpoelstraandKelder:1984}; \opencite{Mercier:1986a}, \citeyear{Mercier:1986b}).}

\subsection{NHR Observations on 13 December 2007}
\label{sec:observations}

We examined NRH radio observations from 8 to 13 December 2007 for its complete observing time. 
Throughout this period, we found slightly enhanced emission over AR~10978 lasting a few hours and, occasionally, other sources at different locations on the disc. On 13 December, the enhancements over AR~10978 were continuously present during the full NRH observing period 
(9:00--15:00~UT). These enhancements can be considered as the signatures of noise storms. We interpret the enhanced emission over the AR as being due to electrons accelerated by reconnection at the QSLs shown 
in Figure 7 
of \inlinecite{Demoulin13}. These accelerated electrons travel along the closed (reconnected) loops anchored at both of the AR borders. Consistent with this interpretation, the intensity isocontour, at 80\,\% of its maximum, in the highest NRH frequency clearly encloses AR 10978 (this has been added in blue continuous line to \fig{nrh2}).
A similar origin can be attributed to the 150.9 MHz isocontour (added in red continuous line to the same figure).
Notice that, taking into account the projection effect, this lowest-frequency contour is well aligned with the top of a loop arcade at the South-West of the AR, which is anchored at both of its borders.

However in this work we are not interested in the radio emission located directly over AR 10978, but
in its vicinity, as we seek evidence of the two-step reconnection process proposed to explain how the AR plasma 
can gain access to the SW (\sect{escape}). Futhermore, as a byproduct of our topological analysis, we have also found null points not linked to EIS upflows (\sect{nulls-closed}). We also investigate whether magnetic reconnection at these external nulls could be the cause of the sporadic bursts. Therefore, we look
for sources of sporadic emission at 150.9~MHz, which is the lowest NRH frequency 
and the one which most likely shows signatures of Type I bursts. We analysed the brightness 
distribution over the solar disc on 13 December with a temporal cadence of 
one second, choosing the time interval for which the ionospheric effects would be
minimal (12:30\,--\,13:40~UT). We verified that this was the case by looking at the East--West 
and North--South one-dimensional brightness distribution diagrams (\url{http://www.secchirh.obspm.fr/index.php}) at 150 and 432~MHz for this day.  In addition to the enhanced emission 
over the AR we observed the existence of regions where the emission fluctuates, increasing and decreasing recurrently, as shown in \fig{nrh1}.
To determine the appearance of these sporadic sources in a more systematic way, we 
used the specific software developed for NRH data analysis that is included in  the Solar-Soft package. We scanned NRH images at 150.9~MHz during  limited periods of time, no longer than 15 minutes, using a square window not larger than 0.25 \Rsun. Proceeding in this way, we found sources between 12:55 and 13:10~UT 
and between 13:25 and 13:40~UT. Panels a\,--\,c in \fig{nrh1} illustrate the brightness distribution at three different hours during the first time interval, while panels d\,--\,f correspond to the second one. 
The sources are not static, as they appear at a slightly different location in each image. 
The white crosses in \fig{nrh1}b indicate the average locations  
of the maximum intensity of the two main sporadic sources found in the first time interval (12:55\,--\,13:10~UT).  
The size of the cross arms indicates the statistical dispersion (1 $\sigma$) in the position estimation. 
The black cross in \fig{nrh1}d illustrates a similar result for the second time interval (13:25\,--\,13:40~UT). 
We have chosen to represent the variable position of the sporadic sources by their averaged location 
in each time interval (white and black crosses in \fig{nrh1}) because we consider it to be the most 
convenient way for comparison with a global and static coronal-field model (see next paragraph). 

Furthermore, to make sure that the observed sporadic sources are not of ionospheric origin,  
we plotted their intensity {\it vs.} time (not shown) and we compared these 
intensity curves with the standard deviation of the continuum corresponding to each of the two 
time intervals. The standard deviations were taken as an upper limit of the ionospheric 
influence; we observe that their values were, at least, an order of 
magnitude lower than the typical intensity of any of the sporadic-source curves. 
Therefore, we conclude that the observed bursty sources have a true solar 
origin.

\subsection{Signatures of the Two-step Reconnection Process}
\label{sect:two-step}

To understand the origin of the sporadic emission in 150.9 MHz during the two time intervals 
discussed in the previous paragraphs, we compare their average locations (white and black crosses in \fig{nrh1}) with field lines computed using the PFSS model described in \sect{model-topo}. This is illustrated in 
\fig{nrh2}. Comparing this figure to \fig{nrh1}, it is clear that the bursty-emission sources are associated with
closed magnetic-field lines. Indeed, a visual comparison of \figs{nrh1}{nrh2} 
reveals that the location of the sporadic sources is distributed along and within the arcade of the large-scale closed loops that shape the streamer belt. 
These bursts may originate from magnetic reconnection between
field lines anchored at the periphery of AR 10978 and closed-field lines with one end connected to 
the quiet Sun at the East of the AR and the other one anchored far to the North. In this case, reconnection is presumably forced by
the magnetic-field dispersion of AR 10978 at successively different locations.
That is why we observe several sources and their locations vary in a dynamic way.
Therefore, we conclude that these sporadic sources can be considered as the coronal manifestation of the first step in the proposed reconnection process (\fig{sketch}a). \fig{nrh2} also includes a set of field lines computed in the neighbourhood of null point N3 (see \fig{closed-nulls}), which have been selected with a higher spatial density and are drawn in light-blue colour to stand over the rest of the field lines. The spatial closeness between this set of field lines and the black and white crosses suggests that, eventually, some of the observed sporadic sources could be the signature of energetic electrons accelerated by magnetic reconnection at null point N3. The location where the emission is observed will, in particular, depend on the ambient-plasma density. Furthermore, if the energy released in the process were not high enough, the emission may not be detectable all along the reconnected loops.  

Finally, our previous analysis indicates that, at the lowest NRH frequency, we observe no radio signature of the second step in the reconnection process sketched in \fig{sketch}b. 
The noise storms we found in the vicinity of AR 10978 are related to closed-loop configurations, as was shown in several other examples (\opencite{Stewart:1977}; \opencite{Svestka:1983}; \opencite{GnezdilovandFomichev:1987}; \opencite{Langetal:1988}; \opencite{Bohme:1989}; \opencite{Kruckeretal:1995}; \opencite{Iwaietal:2012}). 
This  contrasts somewhat with the suggestion by \inlinecite{DelZanna11}, who observed a low-frequency Type III noise-storm jointly with a metric noise-storm and associated both emissions to open-field line 
configurations. 
We have also looked for evidence of Type III bursts, which are known to be due to electrons 
accelerated outwards along the open coronal-field lines (see for example, \opencite{SuzukiandDulk:1985}; 
\opencite{Bastianetal:1998}; \opencite{PickandVilmer:2008}). 
Type III bursts were observed neither by the {\it Nan\c cay Decameter Array} (20\,--\,70~MHz) 
nor by the {\it Waves} experiment (40 kHz\,--\,14 MHz) onboard the {\it Wind} spacecraft. This indicates that 
we have found no evidence of electron acceleration originating at null N1 in the open coronal field.
However, we have to take into account the fact that, despite the large dynamic range
of the {\it Nan\c cay Decameter Array} (greater than 60 dB, \cite{Lecacheux:2000}), 
the presence of a strong source such as an active region can prevent 
the detection of weak Type III bursts.

\section{Summary and Implications for the Slow Solar Wind}
\label{sec:summary}

After the launch of {\it Hinode}, upflows, lasting for several days, were observed at the borders of ARs,
both with XRT and EIT. These observations posed questions about the possible contribution of the upflowing plasma to the slow SW. Several works tackled the problem using different approaches, either by computing the magnetic-field topology in order to explain their origin and the way they could reach the SW, or by comparing their composition to that of the slow SW (see references in \sect{intro}). The link was found by
\inlinecite{vanDriel-Gesztelyi12} through a combined analysis of coronal and {\it in-situ} observations, together with local and global magnetic-field modelling and topology computation. These authors analysed EIS observations of a region crossing the solar disc during CR 2065. During this rotation, AR 10980, the recurrence of AR 10978 which we study in this article, and a small spotless bipolar region formed a quadrupolar large-scale topology. It was proposed that magnetic reconnection at a high-altitude null point, found in the configuration, directed part of the upflowing plasma into the slow SW. During CR 2064, AR 10978, where prominent EIS upflows are observed, is an isolated, simple bipolar region which is completely enclosed by the large-scale loops of the streamer belt 
(see \sect{environment}). In passing over the AR,
the projection of the HCS, determined from a GONG PFSS model, roughly aligns with the AR main polarity inversion line leaving its negative polarity to its western side.   

Searching for the way in which the plasma in EIS upflows, at both sides of AR 10978, can be injected into the SW, we  performed a detailed topological analysis of the region and its surroundings. From our computations, based on a PFSS model, we identify four high-altitude magnetic null-points (\sect{nulls-comp}) located at heights of 25 Mm or greater above the photosphere and 
within $20\degree$ in longitude from the region. The AR polarities are directly linked to only two of these null points (labelled as N3 and N4 in \figs{null-loc}{closed-nulls}). These two null points are associated with closed-field configurations. A third null point (labelled as N2 in \figs{null-loc}{closed-nulls}) is located far North and not directly connected to the AR. This null point also presents a closed field configuration. Therefore, magnetic reconnection at these nulls cannot provide the upflowing plasma a path into the slow SW (\sect{nulls-closed}). 

The fourth null point (labelled as N1 in \figs{null-loc}{open-null}), located to the North-West of AR 10978 and at a height of $\approx$ 108 Mm, is the only one linked to open field lines. A magnetic-interchange reconnection process could occur at this null point (as illustrated in \fig{null1-recon}) and, in this way, the coronal plasma could be channeled into the SW. However, field lines computed from this null-point neighbourhood are not directly connected to AR 10978. This means that, in order to bring the plasma in EIS upflows into the open coronal field, we need a reconnection process working in more than one step. 
In a first step, the AR expansion forces reconnection at QSLs between the AR closed loops 
at its borders and large-scale network fields anchored to the East of the region. As a 
result of this process, illustrated in \fig{sketch}a, plasma in the EIS upflows, mainly from
the eastern side of the AR but also from the western side, is injected 
into large-scale loops anchored close to the following polarity of AR 10978. In a second step, 
these large-scale loops can undergo interchange reconnection with open-field lines 
from the northern CH, as shown in \figs{null1-recon}{sketch}b. The reconnected open-field lines, resulting
from this process, are
anchored to the negative network field on the Sun's surface at one end and reach the source
surface at the other end. The two-step process brings the upflowing plasma into the
open field and, at the same time, displaces the solar connectivity of the open field towards the
solar Equator and, thus, towards the Ecliptic at the source surface.

The results from our topological analysis imply that plasma with typical AR composition should be found in the
negative SW sector in front of the HCS at 1 AU.
This is, indeed, what \inlinecite{Culhane14} found when studying \insitu\ ACE observations at 1 AU.  
Using a back-mapping technique and analysing three characteristics of the SW plasma, measured before the HCS crossing, these authors were able to associate the origin of the SW plasma to AR 10978. Firstly, the C$^{6+}$/C$^{5+}$ and O$^{7+}$/O$^{6+}$ ratios were found to be enhanced, which is a signature of a higher frozen-in temperature origin in the corona. 
Secondly, they determined that in the time period when AR 10978 was crossing the CM, the FIP-bias for the eastern upflow was in the range 3.5 to 4.0 (characteristic of closed AR structures), while the Fe/O ratio measured by ACE also increased by a factor of three to four just before the HCS crossing. Finally, it was found that in the same period the He density was depleted compared to the proton density, which is consistent with the passage of ACE across the streamer belt.  
Therefore, the slow-SW plasma observed by ACE before the HCS crossing seems to have a significant contribution from AR 10978 upflowing plasma. Furthermore, these authors showed that ACE detection of the previous plasma characteristics began five days after AR 10978 reached CM. The time interval for this detection, considering typical slow-SW velocity values, should be around three days; in view of this time difference \inlinecite{Culhane14} concluded that the longer elapsed time could be explained by the addition of the plasma travel time from AR 10978 to null N1, which supports the two-step reconnection process described in the previous paragraph.

The reconnection processes, discussed above, occur in regions where the 
mag\-netic-field intensity is low. Therefore, to find coronal signatures of the energy released, we analysed the radio emission from NRH at frequencies in the range 150\,--\,432 MHz. We found a set of sporadic sources located along and within the large-scale arcade of closed loops that form the streamer belt. These dynamically varying sources are identified during two different time intervals. Their locations (\figs{nrh1}{nrh2}) and evolution suggest that they may originate from magnetic reconnection between field lines anchored at the borders of AR 10978 and closed-field lines with footpoints in quiet-Sun regions to the East of the AR and far to its North. 
This reconnection process is presumably forced by
the magnetic-field dispersion of AR 10978 at successively different locations. We interpret the presence of these sporadic sources as the coronal manifestation of the first step of the proposed reconnection process (\fig{sketch}a) with, eventually, a contribution to energy release at null point N3.  
However, we found no radio-emission evidence of the second-step reconnection process that should occur at null N1. This second step is supported by the presence of plasma with AR composition in the SW at 1 AU.    

For the particular case of AR 10978, our results show how plasma from the eastern and western EIS upflows could be observed {\it in-situ} by ACE. This requires an indirect process that involves magnetic reconnection in at least two steps. Our analysis extends that of \inlinecite{vanDriel-Gesztelyi12}, who showed that part of the AR upflows could become outflows {\it via} direct interchange magnetic reconnection at a high-altitude magnetic null point. Our results, based on observational evidence and the topological properties of the magnetic-field configuration, demonstrate that even when it appears highly improbable for the AR plasma to reach the SW, indirect channels involving a sequence of reconnections at QSLs and high-altitude null points {may} make it possible. Furthermore, to understand the full process, studies such as the one presented here need to be complemented by a detailed analysis of several parameters observed {\it in-situ} (\ie plasma velocity, density, composition, and magnetic field) together with the determination of the outflowing-plasma composition.

%
\begin{acks}
CHM and GDC acknowledge financial support from the Argentinean grants PICT 2007-1790 (ANPCyT), UBACyT 20020100100733, and PIP 2009-100766 (CONICET). The research leading to these results has received funding from the European Commission's Seventh Framework Programme under the grant agreement No. 284461 (eHEROES project). LvDG and DB acknowledge support by STFC Consolidated Grant ST/H00260/1.  LvDG's work was supported by the Hungarian Research grant OTKA K-081421. CHM, AMV, and GDC are members of the Carrera del Investigador Cient\'i fico (CONICET). FAN is a fellow of CONICET.  We 
thank D.M. Long for the enhanced XRT image and H. Morgan for the XRT processing software.  

\end{acks}

 \bibliographystyle{spr-mp-sola-cnd}
 \bibliography{dec-rotation-sw}  

\end{article} 
\end{document}